\documentstyle[preprint,aps,epsfig]{revtex}

\begin{document}
\preprint{gr-qc/9808056}
%must uncomment below...
%\draft
\baselineskip = 1.6\baselineskip
%%%%%%%%%%%%%%%%%%%%%%%%%%%%%%%%%%%%%%%%
\title{
Quantum field and uniformly accelerated oscillator
	}
\author{Hyeong-Chan Kim \thanks{me@taegeug.skku.ac.kr},
	  }

%must uncomment....
\address{
Department of Physics and Institute of Basic Research, 
Sung Kyun Kwan University,\\ Suwon, 440-746,  
Korea}
\date{August 20, 1998}

\maketitle
\begin{abstract}
We present an exact treatment of the influences on a quantum 
scalar field in its Minkowski vacuum state induced by coupling 
of the field to a uniformly accelerated harmonic oscillator. 
We show that there are no radiation from the oscillator 
in the point of view of a uniformly accelerating observer. 
On the other hand, there are radiations in the point of view
of an inertial observer. It is shown that Einstein-Podolsky-Rosen 
(EPR) like correlations of Rindler particles in Minkowski vacuum  
states are modified by a phase factor in front of  the
momentum-symmetric Rindler operators. The exact quantization 
of a time-dependent oscillator coupled to a massless scalar 
field was given.
\end{abstract}
%must uncomment following after finish the paper..
\vspace{.5cm}
\begin{flushleft}
\indent\indent\hspace{4mm}  Keywords: 
	Particle detector model, uniform acceleration, Rindler space.  
		\\
\indent\indent\hspace{4mm}  Pacs number: 04.62.+v
\end{flushleft}

% body of paper here
%\narrowtext
\newpage
%%%%   Model Section   %%%%

\section{Introduction}

It is well known that the Minkowski vacuum state 
has spacelike pair-wise [Einstein-Podolsky-Rosen (EPR) type]
correlations of Rindler particles between the
left and the right Rindler wedge, which
makes the Minkowski vacuum to be described by a 
thermal bath of Rindler particles~\cite{unruh76}. 
In this respect, it is understood that a uniformly accelerating
detector detects thermal radiation of the field.
This was analyzed in more detail by Unruh and 
Wald~\cite{unruhWald} who discussed several apparently 
paradoxical aspects related to causality, energy considerations, 
and particle creation from the detector. The conclusion for the 
presence of the radiation by Unruh and Wald was criticized by
Grove~\cite{grove} who claimed that the appearance of particle 
is not a dynamical process but an artifact of the state reduction. 
Massar, Parentani, and  Brout \cite{massar},and 
Hinterleitner~\cite{hin} resume the discussions by use of a 
solvable model presented by Raine, Sciama, and
Grove~\cite{raine}:
In the model of a harmonic oscillator coupled to a scalar quantum
field they investigated the mean value of the energy-momentum
tensor of the field.  They concluded that a uniformly
accelrated oscillator does not radiate.

This contradiction was discussed by Audretsch and 
M\"{u}ller~\cite{aud} using two level detector. They pointed 
out the difference between the stress tensor and the particle
number and then stressed the importance of the built-up 
correlations and the quantum measurement process.
Lau~\cite{lau} also discussed the radiation from the 
oscillator and the disappearence of the EPR- correlations. 
However, in these literatures, the authors assume the 
coupling of the detector to the quantum field is weak.
So, the back reaction to the field was not counted exactly,
and the mechanism of particle creation is unclear. 
This is the reason why the previous literatures
fail to provide all of non-perturbative aspects.
In this paper, we focus on obtaining the exact vacuum to vacuum 
relations.  We summarize 
the main results in this paper in Fig. (1), 
which describes the relations between several different vacua. 
\begin{figure*}[bth]
\vspace{0.5cm}
\centerline{
\epsfig{figure=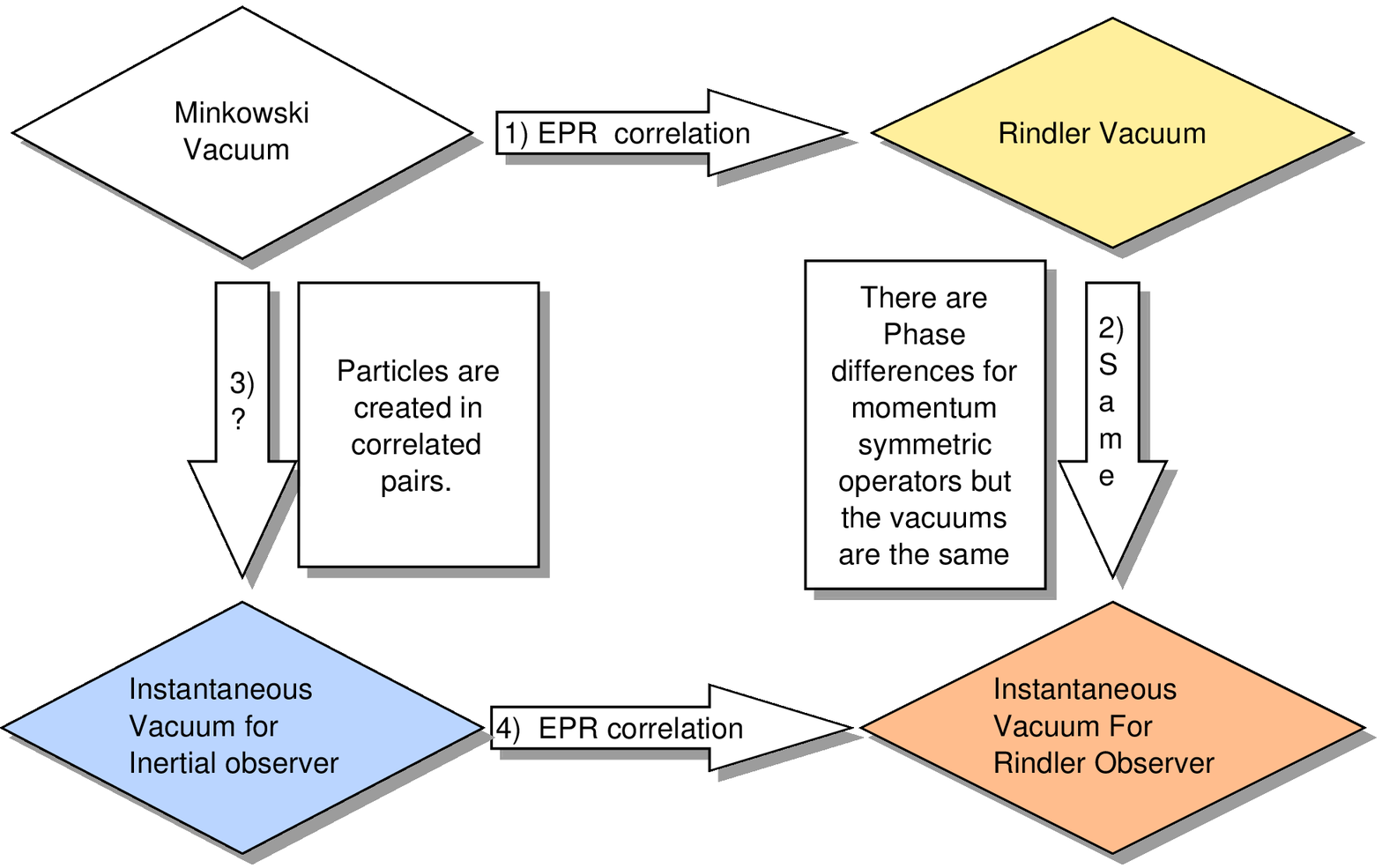,height=8cm}}
{\footnotesize Fig. 1.  Vacuum structure for each observer
	with and without detector \\
1), 4) The Minkowski vacuum state is
a coherent state of Bogoliubov-Bardeen-Cooper-Schrieffer type,
in which particles are pairwise correlated (a particle in the positive
Rindler wedge are correlated with a particle in the left Rindler
wedge.)~\cite{unruh76,takagi} \\
2) The Rindler vacuum is the same with the instantaneous
vacuum for a uniformly accelerating observer. There are only
some phase shift for momentum symmetric states with non-zero
number of particles. \\
3) The phase shift in 2) makes crucial differences for the vacuum
structures between Minkowski and the instantaneous
observers. There are net creation of particles from the oscillator. 
 }
\end{figure*}
The relation 1) is well known.  
In the present paper, we clarify the other three 
relations in exact forms which need to calculate
the back reaction of the oscillator to the field. 
In due course, we define each ground state from
the creation and annihilation operators
and find the relations between each sets of operators.
During these calculations, we succeed in quantizing the complete 
system with a time-dependent coupling constant, which 
is given as an appendices. 

The outline of this  work is as follows. 
In Sec. II, we introduce our notation and briefly
sketch our calculational procedure.  
Section III is devoted to compute the particle 
creation of from the oscillator and the EPR-correlations
of the Minkowski vacuum. 
In Sec. IV, we summarize and discuss the result.  
There are two appendices which are devoted 
(A) to develope an exact quantization scheme 
for the coupled system in the respect of Rindler observer
and (B) to get the formal relations between the 
operators at initial and final time in Rindler space.

%%%%%%%%%%%%%%%%%%%%%%%%%%%%%%%%%%
%
%
\section{Definitions, notations, and sketch of procedure}

The (1+1) dimensional Minkowski space can be devided 
into 4 regions in the respect of a Rindler observers.
Let us call each regions by Future (F), Past (P), 
Right~(R, $\sigma=+$),
and Left~(L, $\sigma=-$) quadrant of Rindler space. 
We parametrize the region R using 
$(\tau,\rho)$:
\begin{eqnarray}
x = \frac{1}{a} ~e^{a\rho} \cosh a\tau, 
~~ t =  \frac{1}{a} ~e^{a\rho} \sinh a \tau,
\end{eqnarray}
where $(t,x)$ is the Minkowski coordinate basis. L also can 
be parametrized by similar cordinates patch, but we do not 
write it explicitly in this paper. 

The free field in (1+1) dimension is given by
\begin{eqnarray}
\label{phi:ab}
\phi^0(t,x)&=& \int dk [ a_k U_k(t,x) + a_k^\dagger U_k^*(t,x)] 
	~\mbox{in Minkowski space,} \\
	&=& \sum_\sigma \int d\kappa [b^{(\sigma)}_\kappa 
		u_\kappa^{(\sigma)} (\tau,\rho)
		+ b^{(\sigma) \dagger}_\kappa  u_\kappa^{(\sigma)*}
		(\tau,\rho)] ~
				\mbox{in Rindler space}, \nonumber 
\end{eqnarray}
where the mode solutions are
 $  \displaystyle 
u_\kappa^{(\sigma)}= \frac{1}{\sqrt{4 \pi \Omega}} 
	e^{-i \sigma \Omega \tau +i \kappa \rho }$ and
$\displaystyle  U_k =\frac{1}{\sqrt{4\pi \omega}}e^{-i \omega t+ik x}$. 
From now on we represent $\omega=|k|$, and $\Omega=|\kappa|$.
The relation between $a_k, ~a_k^\dagger$ and 
$b^{(\sigma)}_\kappa ,~b^{(\sigma)\dagger}_\kappa$ is given by 
Bogoliubov transformation
\begin{eqnarray}
%\label{b:a}
b_\kappa^{(\sigma)}= \int dk \left( \alpha^{(\sigma)}_{\kappa k} a_k
	+\beta^{(\sigma)}_{\kappa k} a_k^\dagger\right) ,
\end{eqnarray}
which is simplified to be 
\begin{eqnarray}
\label{b:d}
b_\kappa^{(\sigma)} = [1+ N(\pi \Omega/a) ]^{1/2}d^{(\sigma)}_\kappa 
	+N^{1/2}(\pi \Omega/a) d_\kappa^{(-\sigma)\dagger} .
\end{eqnarray}
Where $\displaystyle N(\alpha) =\frac{1}{e^{2\alpha}-1} $ 
and $\displaystyle d_\kappa^{(\sigma)} = \int_{-\infty}^\infty 
dk  p_\kappa^{(\sigma)}(k) a_k  $. 
In Eq. (\ref{b:d}), the function
\begin{eqnarray}
p_\kappa^{(\sigma)}(k) &=& \frac{1}{\sqrt{4 \pi a \omega } } 
\left(\frac{a}{\omega} \right)^{-i \sigma \kappa/a} 
	~\mbox{for} ~ \kappa k > 0, \\
&=& 0 ~~ ~\mbox{for}~ \kappa k \leq 0  \nonumber
\end{eqnarray} 
is complete and orthonormal in the following sense 
(see Ref. \cite{takagi} for higher
dimensions):
\begin{eqnarray}
\label{p}
\int_{-\infty}^{\infty}dk p_\kappa^{(\sigma)*}(k) 
p_{\kappa'}^{(\sigma')} (k) &=& \delta_{\sigma \sigma'}
	\delta(\kappa-	\kappa') , \\
\sum_\sigma \int d\kappa p_\kappa^{(\sigma)*}(k)
	p_\kappa^{(\sigma)}(k') &=& \delta(k-k')  .
\end{eqnarray}
Therefore, $d_\kappa^{(\sigma)\dagger}$ 
creates and 
$d_\kappa^{(\sigma)}$  annihilates a Minkowski particle
and satisfies  
\begin{eqnarray}
%\label{}
[d_{\kappa}^{(\sigma)}, d_{\kappa' }^{(\sigma')\dagger}]
	=\delta_{\sigma,\sigma'} \delta(\kappa-\kappa') .
\end{eqnarray}
Minkowski vacuum is defined by
\begin{eqnarray}
\label{0:M1}
d_{\kappa}^{(\sigma)} |0\rangle_M =0 
	,  ~~\sigma=\pm ,
\end{eqnarray}
and Rindler vacuum also is defined by
\begin{eqnarray}
%\label{}
b_\kappa^{(\sigma)} |0\rangle_R =0.
\end{eqnarray} 

By solving Eq. (\ref{0:M1}) using Eq. (\ref{b:d}) we get 
the well known non-local pair-wise correlation of the 
Minkowski vacuum with respect to an accelerated 
observer, which can be summarized by the following formulae:
\begin{eqnarray}
\label{0M:0R}
|0\rangle_M = \prod_\kappa (1-r^2)^{1/2}
		 \exp(r b_\kappa^{(\sigma)\dagger}
		 b_\kappa^{(-\sigma) \dagger}  ) |0 \rangle_R ,
\end{eqnarray}
where 
\begin{eqnarray}
\label{r}
 r= e^{-\pi \Omega/a}.
\end{eqnarray}
Up to this point, we follow the beautiful review of Takagi~\cite{takagi}.
In fact, we rederived these results because his calculation
for $p_\kappa^{(\sigma)}(k)$ cannot be directly applicable
to $1+1$ dimension. 

Now let us consider a detector of oscillator $q(\tau)$ which is 
minimally coupled to a massless real scalar 
field $\phi(t,x)$ in two dimensions.
 This model is already discussed in Ref. \cite{hckim},
in which authors calculated the radiation from the oscillator 
with a time-varing coupling constant.
The only difference of the model from other 
Refs.~\cite{hin,massar,unruhZurek} 
are the time dependency of the coupling constant.
The action of the system is
\begin{eqnarray}\label{ac}
S &=& \int \mbox{d}x \mbox{d}t
    \frac{1}{2}\left\{ \left(\frac{\partial}{\partial t} 
	\phi(t,x) \right)^2 
	- \left( \frac{\partial}{\partial x} \phi(t,x)\right)^2
	\right\}  \\
  &+& \int d\rho d\tau  \left\{ \frac{1}{2} m 
		\left(\frac{d q(\tau)}{d \tau} \right)^2 
	-\frac{1}{2} m \omega_0^2 q^2(\tau) -e(\tau)q(\tau) 
	\frac{ d\phi}{d \tau} \left(t,x\right) \right\} \delta(\rho)
       . \nonumber
\end{eqnarray}
The oscillator follows the uniformly accelerated trajectory
$(\rho=0)$ in R.

We decompose the field using mode solutions
even in the presence of the oscillator. The mode solution
outside the oscillator satisfies the free field equation. 
So we continually use $U_k$ and $u^{(\sigma)}_\kappa$
as the mode solutions in each regions. 
The instantaneous creation and 
annihilation operators depend on time and the
mode expansion of the field is 
\begin{eqnarray}
\label{phi:AB}
\phi(t,x)&=& \int dk [ A_k(t) U_k(t,x) + A_k^\dagger(t) U_k^*(t,x)] 
		~\mbox{in Minkowski space} , \\
	&=& \sum_\sigma \int d\kappa [B^{(\sigma)}_\kappa(\tau) 
		u_\kappa^{(\sigma)}(\tau,\rho)
		+ B^{(\sigma) \dagger}_\kappa(\tau)
		  u_\kappa^{(\sigma)*}(\tau,\rho)] ~
				\mbox{in Rindler space}. \nonumber 
\end{eqnarray}
The position operator of the oscillator is described by
\begin{eqnarray}
\label{q:a}
q(\tau) = \frac{ a(\tau) + a^\dagger(\tau) }{\sqrt{2 m \omega_0}} ,
\end{eqnarray}
where $a(\tau),~a^\dagger(\tau) $ are the time-dependent 
ladder operators of the oscillator.

We are to consider a system which evolves from the initial state 
that the scalar field and the oscillator are completely
decoupled. Therefore, the vacuum $|0\rangle_M$ for initial state is 
is described by the direct product of the Minkowski vacuum
and the ground state of the oscillator at $\tau_0$:
\begin{eqnarray}
\label{0:M}
a_k|0\rangle_M =0 = a(\tau_0)|0\rangle_M.
\end{eqnarray}
Where we use the same notation as the vacuum state
without oscillator.
Similarily, we define the initial Rindler vacuum state
$\left|0\right>_R$ as
\begin{eqnarray} \label{0:b}
b^{(\sigma)}_\kappa \left|0\right>_R=0 = a(\tau_0)\left|0 \right>_R .
\end{eqnarray}

The Fock space at  time $t$ of an inertial observer
should be defined by the instantaneous eigenmodes. 
From~(\ref{phi:AB})  we define the instantaneous
vacuum state $|0\rangle_A$ to be
\begin{eqnarray}
\label{0:A}
A_k(t) |0\rangle_{A}  =0= a(\tau) | 0\rangle_{A} 
	\mbox{ in Minkowski space},
\end{eqnarray}
where $\tau$ is the propertime of the oscillator at the 
hypersurface given by $t=\mbox{const}$. 
Similarily, we define the instantaneous vacuum 
of a uniformly accelerating observer at $\tau$ as 
\begin{eqnarray}
\label{0:B}
B_\kappa^{(\sigma)} (\tau)\left|0\right>_B =0= a(\tau)\left|0\right>_B
	\mbox{ in Rindler space}.
\end{eqnarray}
In this equation $B_\kappa^{(\sigma)}(\tau)$ and 
$B_\kappa^{(\sigma)\dagger}(\tau)$ can be obtained from the
formula:
\begin{eqnarray}\label{B}
B_\kappa^{(\sigma)}(\tau) &=& \int d\rho ~
	u_\kappa^{(\sigma)*} (\tau,\rho) i\Pi_\phi(t,x)
	-\int d\rho  ~\phi(t,x) i\partial_\tau 
	u_\kappa^{(\sigma)*} (\tau,\rho) ,
	  \\
B_\kappa^{(\sigma) \dagger} (\tau) &=&-\int d\rho~
	u_\kappa^{(\sigma)}(\tau,\rho)
	 i\Pi_\phi(t,x)
	+\int d\rho~\phi(t,x) i\partial_t 
	u_\kappa^{(\sigma)} (\tau,\rho) , \nonumber
\end{eqnarray}
where $\Pi_\phi$ is the conjugate momentum of $\phi$.
With this form, $B_\kappa^{(\sigma)}(\tau), ~ 
	B_\kappa^{(\sigma)\dagger}(\tau)$ 
automatically satisfy 
\begin{eqnarray}
\label{comm}
&& \left[B^{(\sigma)}_\kappa(\tau), B_q^{(\sigma')\dagger}(\tau) \right] = 
		\delta_{\sigma\sigma'} \delta(\kappa-q), ~ \\
&& \left[B^{(\sigma)}_\kappa(\tau), a(\tau)\right] =
	 \left[B^{(\sigma)}_\kappa(\tau),
		 a^\dagger(\tau)\right]=0 ,  \nonumber
\end{eqnarray}
which allows a natural definition of Fock space as 
eigenstates of $B_\kappa^{(\sigma)\dagger}B_\kappa^{(\sigma)}$
and $a^\dagger a$.

From Eq. (\ref{phi:AB})  we get the following relations
between the two sets of the annihilation and the 
creation operators:
\begin{eqnarray}
\label{A:B}
A_k &=& \int d\kappa \left[ \{1+N \} ^{1/2} p_\kappa^{(\sigma)^*}(k)
 B_\kappa^{(\sigma)} - N^{1/2} p_\kappa^{(\sigma)*}(k) 
 B_\kappa^{(\sigma) \dagger} \right]  \\
B_\kappa^{(\sigma)} &=& (1+N)^{1/2} D_\kappa^{(\sigma)} +
		N^{1/2} D_\kappa^{(-\sigma) \dagger}  \label{B:D}
\end{eqnarray}
where $D_\kappa^{(\sigma)} = 
	\int dk p_\kappa^{(\sigma)} (k) A_k$.
These relations are exactly the same with Eq. (\ref{b:d}),
which means the 
ground state of $A$ is a EPR-like correlated state of 
Rindler particles created by $B^{(\sigma)\dagger}_\kappa$.

What 
we would like to know is the inter-relations 
between the instantaneous 
operators $B^{(\sigma)}_\kappa,~a(\tau) $, 
$ A_k$ and the initial operators $a_k$, 
$b_\kappa^{(\sigma)}$, $a(\tau_0)$.  
Note that $B^{(\sigma)}_\kappa,~ A_k,$ and $a(\tau)$ depend
on time but
$a_k$ and $b^{(\sigma)}_\kappa$ are independent of time. 
So  we set the initial value of  $B^{(\sigma)}_\kappa,~ A_k$ as
$b_\kappa^{(\sigma)}$, $a_k$. 
%At this point, we need one
%more operators $B_Q(\tau)$ which describes 
%the initial state of the oscillator. 
The time evolution of the system is completely determined by
the relations between these two sets of operators
[$A_k, B_\kappa^{(\sigma)}, a(\tau); a_k, b_\kappa^{(\sigma)}, 
a(\tau_0)$] in appendix A. 
In appendix B, we calculate the asymptotic limit of the system
by taking $\tau_0\rightarrow -\infty$ and $e(\tau)=e$.
In fact, we only use the final result of appendix B~(\ref{B:b}),
which relates the in and out creation and annihilation operators
in Rindler frame. The details which leads to (\ref{B:b}) is not 
relevant for later discussions. 

For later convinence, we define the momentum symmetric and 
asymmetric creation and annihilation operators 
\begin{eqnarray}
\label{apm:a+-}
a_{\omega\pm}
	&=& \frac{a_\omega \pm 
	a_{-\omega}}{\sqrt{2}}  ,\\
b_{\Omega\pm}^{(\sigma)} &=&
	\frac{ b_\Omega^{(\sigma)}
 \pm b_{-\Omega}^{(\sigma)} }{\sqrt{2}} . \nonumber
\end{eqnarray}
We use similar notations for other operators.

%%%%%%%%%%%%%%%%%%%%%%%%%%%%%%%%%
%  
%
%
%
%%%%%%%%%%%%%%%%%%%%%%%%%%%%%%%%%
\section{Particle creation by accelerating oscillator}

We now consider the system in the asymptotic limit 
$\tau_0 \rightarrow -\infty$. 
We set the initial state at~$\tau = \tau_0$ to be the direct 
product of Minkowski vacuum and the $n^{\rm th}$ 
state of the oscillator.  
In this respect, we regard the instantaneous ground 
state~(\ref{0:A}) and ~(\ref{0:B}) as the out vacuum
state for the inertial observers and the uniformly accelerating 
observers each other.

%%%%%%%%%%%%%%%%%%%%%%%
\subsection{Uniformly accelerating observer point of view}

At first, let us consider the system in the point of view
of the uniformly accelerating observers.
The transformation (\ref{B:b}) is most simple when we 
use the momentum symmetric and asymmetric creation 
and annihilation operators: 
\begin{eqnarray}
\label{B+:b+}
B^{(+)}_{\Omega +}(\tau) =
	P(\Omega) b^{(+)}_{\Omega +},
~ B^{(+)}_{\Omega  -}(\tau) = b^{(+)}_{\Omega -}, ~
B^{(-)}_{\Omega\pm}(\tau)= b^{(-)}_{\Omega \pm} ,
\end{eqnarray}
where $\displaystyle P(\Omega)
= \frac{\chi(\Omega)}{\chi^*(\Omega)}$.
The  interaction modifies only the phase of the symmetric 
operators. In the case of a creation operator of a
momentum eigenstate, there occurs relative phase 
shift due to the phase factor 
$P $ 
in Eq. (\ref{B+:b+}). The maximal phase shift occurs
at $\Omega=\omega_0$, which is the resonance frequency 
of the oscillator.In Eq.~(\ref{B+:b+}), there  
is no mixing of the creation and the annihilation operators. 
Therefore, the two ground states in Eqs.~(\ref{0:b}) and (\ref{0:B})
are the same for the scalar field.   

From Eqs. (\ref{B:D}) and (\ref{B+:b+}) we get
\begin{eqnarray}
\label{B:d}
P(\Omega) B_{\Omega+}^{(+)} &=&
	(1+N)^{1/2}d_{\Omega+}^{(+)} 
	+ N^{1/2} d_{\Omega+}^{(-)\dagger}  ,\\
 B_{\Omega+}^{(-)} &=&
	(1+N)^{1/2}d_{\Omega+}^{(-)} 
	+ N^{1/2} d_{\Omega+}^{(+)\dagger} . \nonumber
\end{eqnarray}
By solving Eq. (\ref{0:M}) using (\ref{B:d}) 
 we get the relation between the
initial state and the out vacuum state:
\begin{eqnarray}
\label{0M:0B}
|0\rangle_M= \prod_\Omega (1-r^2) \exp \left[ r\left(
	P(\Omega)
	B_{\Omega+}^{(+)\dagger} B_{\Omega+}^{(-)\dagger} 
	+ B_{\Omega-}^{(+)\dagger} B_{\Omega-}^{(-)\dagger}
\right) \right] |0\rangle_B, 
\end{eqnarray}
where $r$ is given by Eq. (\ref{r}).
The oscillator does not influence on 
the field but is determined completely by the symmetric
mode of the field [See  Eq. (\ref{a:b})].
The number of Rindler particles in the initial state (\ref{0M:0B})
is 
\begin{eqnarray}
\label{BB}
\left\langle B^{(+)\dagger}_\kappa B^{(+)}_q \right\rangle_M
=  N\left(\frac{\Omega}{a}\right)
	% \left\{1+ 8 m^2 \epsilon^2
	%\Omega^2 \left| \chi\left(\Omega\right) \right|^2
	%\right\} 
	\delta(\kappa - q) = 
	\left\langle b^{(+)\dagger}_\kappa b^{(+)}_q \right\rangle_M
\end{eqnarray}
This result vividly shows there are no additional particle
creation from the oscillator in the point of view of a uniformly
accelerating observer.
We claim stringent constraint for the particle creation that
there are no additional creation  of particle from the oscillator
for every initial states of the oscillator and the field.
The proof for this assertion is simple because
$B^{(+)\dagger}_{\Omega\pm} B^{(+)}_{\Omega'\pm}=
b^{(+)\dagger}_{\Omega\pm} b^{(+)}_{\Omega\pm}$
is an operator level identity for the momentum symmetric 
or asymmetric particles.
On the other hand, the transition amplitude for different
number states of symmetric particles acquires the phase factor. 

%%%%%%%%%%%%%%%%%%%%%%%%%%%5
\subsection{Inertial observer point of view}
Next, let us consider the same system in
the inertial observers point of view.

There are delta function type contribution along the
past event horizon of the oscillator ($t+x=0$) 
but it depends on the initial condition of the oscillator 
and commutes with $B_\kappa^{(\sigma)}$. 
Therefore we ignore it in the following calculations. 
From Eqs. (\ref{B:D}) and (\ref{B:d}), we get the following relations:
\begin{eqnarray}
\label{D:d}
D_{\Omega+}^{(+)} &=& 
	\left[P +N\left(P-1\right) \right] d_{\Omega+}^{(+)}  
	+\sqrt{N(1+N)}  \left( P-1
	\right)d_{\Omega+}^{(-)\dagger} ,\\
D_{\Omega+}^{(-)\dagger} &=& \sqrt{N(1+N)}
	 \left(1-P\right) d_{\Omega+}^{(+)} 
	+\left[1+N\left(1-P\right) 
	   \right]d_{\Omega+}^{(-)\dagger}   .
	  \nonumber
\end{eqnarray}
We solve Eq. (\ref{0:M}) by use of (\ref{D:d}), 
then we get  the instantaneous states  representation 
of the initial state in Minkowski space:
\begin{eqnarray}
\label{0M:0A}
|0\rangle_M=\prod_\Omega (1-|R|^2)^{1/2}\exp\left(R 
	D^{(+)\dagger}_{\Omega+}
	D_{\Omega+}^{(-)\dagger} \right) |0\rangle_A
\end{eqnarray}
where
\begin{eqnarray}
\label{R}
R=\frac{4 i \epsilon m \Omega \chi(\Omega) \sqrt{N(1+N)}}{
	1-4i \epsilon m \Omega \chi(\Omega) N} .
\end{eqnarray}
Its size is
\begin{eqnarray}
%\label{}
|R|^2 =\frac{16m^2 \epsilon^2 \Omega^2|\chi(\Omega)|^2N(N+1)}{
	1+16m^2 \epsilon^2 \Omega^2|\chi(\Omega)|^2N(N+1)} 
	= \frac{4m^2 \epsilon^2 \Omega^2|\chi(\Omega)|^2}{
	\sinh^2(\pi \Omega/a) + 4m^2 \epsilon^2 
	\Omega^2|\chi(\Omega)|^2}.
\end{eqnarray}
The presence of oscillator results in the mixing of the 
Minkowski annihilation and creation operators, so, there are 
net creations of particles from the oscillator.  
The number of particles
created by the oscillator in unit time and space is 
\begin{eqnarray}
\label{N}
{\cal N}= _M\left< 0 \right|  \int d k 
	A_k^{\dagger}
	A_k \left|0\right>_M  &=&
_M\left< 0 \right| \sum_\sigma \int d \Omega 
	D_\Omega^{(\sigma) \dagger}
	D_\Omega^{(\sigma)} \left|0\right>_M  \\
	&=& 8m^2 \epsilon^2 \int_0^\infty d\Omega \Omega^2
		N (N+1) \left|\chi(\Omega)\right|^2 . \nonumber \\
	&=& 2m^2 \epsilon^2 \int_0^\infty d\Omega \Omega^2
	\frac{|\chi(\Omega)|^2}{\sinh^2\left(\pi \Omega/a \right)}
	\nonumber
\end{eqnarray} 
As can be seen in Eq. (\ref{N})  the number of created particles
vanishes for zero acceleration and monotonically increases
according to the acceleration.  
We explicitly integrate the integration in Eq.~(\ref{N}) 
using the contour in Fig. (2) after changing the integration range
$\int_0^\infty$ to $\displaystyle \frac{1}{2}\int_{-\infty}^\infty$. 
\begin{figure*}[bth]
\vspace{0.5cm}
\centerline{
\epsfig{figure=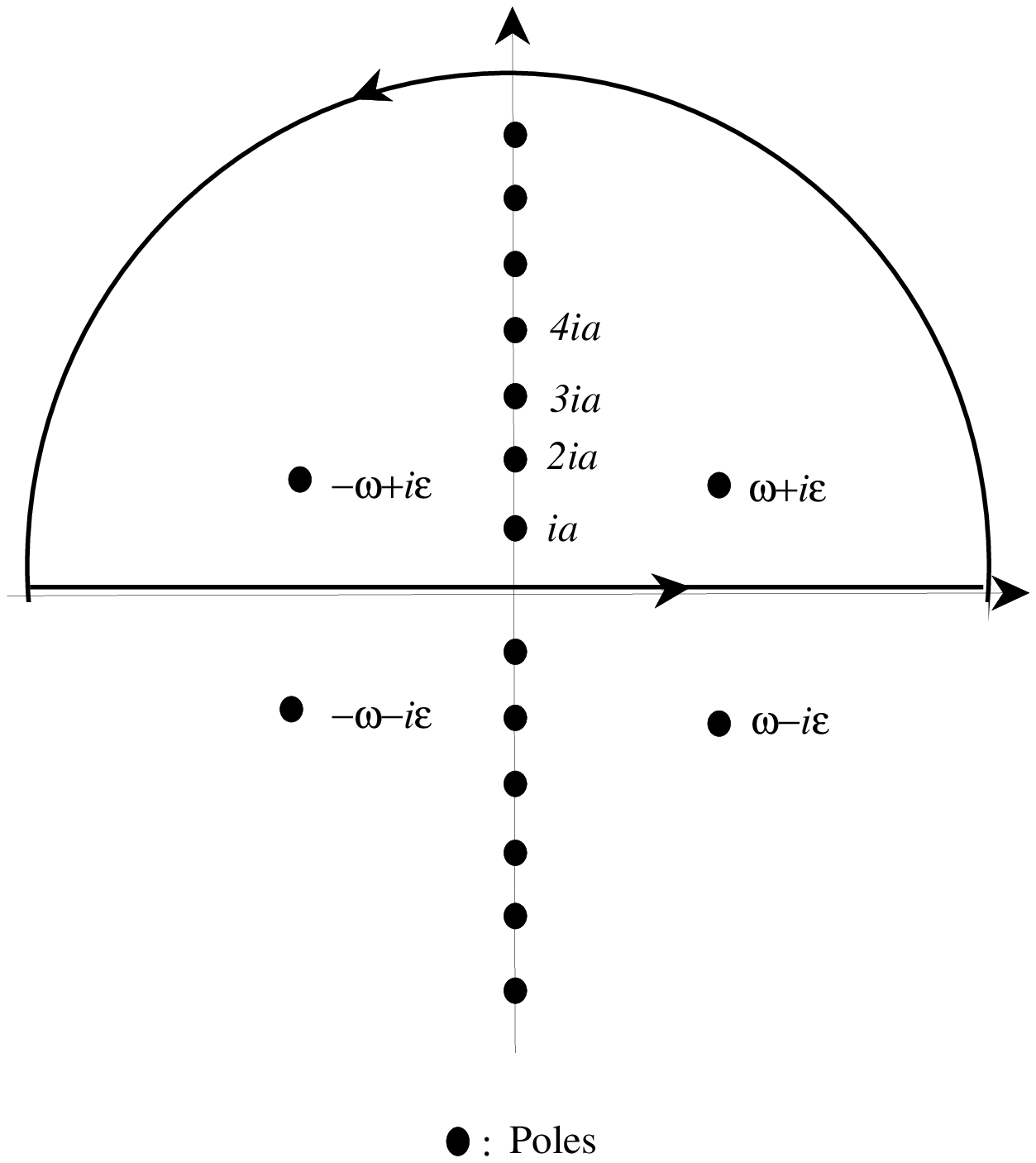,height=8cm}}
{\footnotesize Fig. (2). contour in complex $\Omega$ plane.
There are two kinds of poles which should be considered.
First is poles given by the properties of the oscillator which
is located at $\Omega=\pm\omega_0\pm i\epsilon$ and 
second is the reaction due to acceleration which are located
at $\Omega=ina$ along the imaginary axis. Here $n$ is an
integer.
 }
\end{figure*}
The number of created particles from the  uniformly 
accelerated oscillator is
\begin{eqnarray}
\label{N:sum}
{\cal N}&=& \frac{\pi \epsilon}{\omega_0}m^2 Re\left[ 
	\frac{\omega_0+i\epsilon}{
	\sinh^2\left\{\pi(\omega_0+i\epsilon)/a \right\}} \right]  
	+  \frac{8\epsilon^2a^3}{\pi}\sum_{n=1}^\infty 
	n m^4 |\chi(\Omega)|^4_{\Omega=ina}
	\left[ (na)^4- \omega_0^4 \right] .
\end{eqnarray}
Note that ${\cal N}$ is of order $O(e^2)$ and is monotonically
increasing function of $a$. Each term in 
Eq.~(\ref{N:sum}) is positive definite for $a > 0$.

%%%%%%%%%%%%%%%%%%%%%%%%%%%%%%%%%%%
%
%
%
%
%\section{Oscillator}

%%%%%%%%%%%%%%%%%%%%%%%%%%%%%%%%%%%
%
%
%
%
\section{summary and discussions}

We have obtained the vacuum to vacuum relations
of the a massless scalar field interacting with a 
harmonic oscillator both for the inertial and the uniformly 
accelerating observers. There are four kinds of vacuums
each are related with the inertial observers (M) or the uniformly 
acclerating observers (R) and the in (I) or out~(O) Fock space. 
We set the initial vacuum state to be (MI) the direct product 
of the Minkowski vacuum and the ground state of the oscillator. 

The relation between the MI
and the RO vacuums is modified by the
phase factor attatched at the creation operator of 
Rindler particle ($B^{(+)\dagger}_{\Omega +}$).
The phase factor does not give any noticible change to
physical observables localized in R or L of  Rindler wedge.  
%Therefore, it does not change
%the fact that the Minkowski vacuum state is described by
%the thermal distribution of Rindler particles when viewed 
%by a uniformly accelerating observer.
%Because of the distribution (\ref{0M:0B}), it is quite natural
%that there is no additional particle creations due to the 
%oscillator.
%This fact is very different from the work of Audretsch
%and Muller~\cite{aud} in which they treat two level atom
%in a perturbative manner. 
We present two comments in order:
First, there is no radiation of particles to all order of 
$e$ when viewed by an uniformly accelerating observer. 
Second, the final state of the oscillator is 
%not its ground state, by the way, 
%it does not give any reaction to the field.
completely determined by the 
field [See Eq.~(\ref{a:b})]. 
If we set the acceleration of the oscillator to be zero,
we get the following uncertainty relation of the oscillator:
\begin{eqnarray}
%\label{}
\left<\Delta p \Delta q\right>^2= \frac{\omega_0^2 (\pi-\xi)}{
	4 \pi^2\omega_I^2} \left[ (\pi -\xi)-\frac{2\epsilon\Omega_I}{
	\omega_0^2}\left(\gamma + \ln \frac{\omega_0}{\Gamma}
	\right)   \right] ,
\end{eqnarray}
where $\Gamma$ is the UV cutoff which is introduced in
Ref.~\cite{hckim} and $\gamma$ is Euler constant. 
Note that  the uncertainty go to $\displaystyle 
\frac{1}{2}$ when $\epsilon \rightarrow 0$. This 
is  consistent with Anglin, Laflamme, Zurek, and Paz~\cite{anglin},
where they showed that the final state of the oscillator is 
the ground state if the coupling is extremally small.  

In the point of view of an inertial observer,
 there are radiated particles with its number given by
Eq. (\ref{N:sum}). The number of the radiated particles
is of order $e^2$ for small $e$ and vanishes for $a=0$.  
Therefore, these particles are created by the 
acceleration of the oscillator. 
In fact, there are another particles created by the 
oscillator which 
we neglected in this paper. As pointed out by 
Unruh~\cite{unruh92} there is delta function
like radiation along the past lightcone of the oscillator.
This term only depends on $a(\tau_0), ~ a^\dagger(\tau_0)$ 
which commute with $B_\kappa^{(\sigma)}, $ and
$A_k$. So we can justify the ignorance of it in this paper.
 
Another point to note is that there is symmetries on the 
Minkowski vacuum (\ref{0M:0B})  and (\ref{0M:0A})
for the pair interchanges: 
\begin{eqnarray}
\label{EPR}
\left(B_{\Omega+}^{(\sigma)\dagger}, 
	B_{\Omega-}^{(\sigma)\dagger}\right)  &\leftrightarrow  &
	\left(B_{\Omega+}^{(-\sigma)\dagger},
	B_{\Omega-}^{(-\sigma)\dagger}\right)  , \\
\left(B_{\Omega+}^{(+)\dagger} ,B_{\Omega+}^{(-)\dagger}\right)
	&\leftrightarrow& \left(\frac{\chi^*(\Omega)}{\chi(\Omega)}
	B_{\Omega-}^{(+)\dagger}, B_{\Omega-}^{(-)\dagger}
	 \right)  \label{+:-}
\end{eqnarray}
These are the modified symmetries which represent the 
EPR like correlations of Minkowski vacuum viewed by a 
uniformly accelerated observer.
If the oscillator decouple with the field, 
the  phase factor $\chi^*/\chi$ in Eq. (\ref{+:-}) becomes 1
and Eqs. (\ref{EPR}) and (\ref{+:-}) 
reproduce complete EPR correlation.
%
% Therefore, in our calculation, the
%correlations between the spacelike interval of the Minkowski
%vacuum are not exactly hold. If one consider the
%symmetric states [this is not known before] or asymmetric
%state [see paper...] the correlations conserved exactly
%but for pure momentum states, the correlations
%are broken by the phase factor.

A new non-trivial fact, from Eq. (\ref{0M:0A}), is that there 
remain symmetries 
\begin{eqnarray}
%\label{}
 D_{\Omega_+}^{(\sigma)\dagger}  & \leftrightarrow &
	D_{\Omega_+}^{(-\sigma)\dagger} .
\end{eqnarray}
to an inertial observer. This symmetries leave corrrelations 
between the created Minkowski particles in pair. 

%%%%%%%%%%%%%%%%%%%%%%%%%%%%%%%%%%%%%%%
%
%
%
\newpage
\section{acknowledgements}
This work was supported by the Ministry of Education
(BSRI/97-1419), the KOSEF (Grant No. 95-0702-04-01-3 and 
through CTP, SNU), and Faculty Research Fund, Sungkyunkwan University, 1997.

%%%%%%%%%%%%%%%%%%%%%%%%%%%%%%%%%%%%%%
%
%
%
%
\section{appendix A: Quantization and time evolution of the 
coupled system}

The uniformly accelerated oscillator always stay in R and
its motion is simplest in this coordinates patch. 
Therefore, in this appendix, we only consider the system in R. 
  
Varying the action~(\ref{ac}) with respect to $\phi(t,x)$ 
and $q(\tau)$, we get the Heisenberg equation of motion 
for  the field and the oscillator
\begin{eqnarray}
\Box \phi(t,x) &=& 
   \frac{de(\tau)q(\tau)}{d \tau} \delta (\rho),  \label{2}\\
m \left( \frac{d}{d \tau}\right)^2 q(\tau) 
   &+& m \omega_0^2 q(\tau)
   = - e(\tau)\frac{d \phi}{d \tau}(t(\tau),x(\tau)), \label{3}
\end{eqnarray}
The formal solution to Eq.~(\ref{2}), by using the 
two-dimensional Green's function, is
\begin{eqnarray} \label{phi:uv}
\phi(t,x) = \phi^0(t,x) + \frac{e(\tau_{\rm ret})}{2} q(\tau_{\rm ret}),
\end{eqnarray}
where $\tau_{\rm ret}$ is the value of $\tau$ at the intersection of the
past lightcone of $(t,x)$ and the trajectory of the oscillator.
After the substitution of  the solution (\ref{phi:uv}) 
into Eq.~(\ref{3}) we get
\begin{eqnarray}\label{q'':F}
  \ddot{q}(\tau) + \frac{d\ln M(\tau)}{d \tau}\dot{q}(\tau) 
	+ \omega^2(\tau) q(\tau) = F(\tau) .
\end{eqnarray}
In this equation 
\begin{eqnarray}
M(\tau) &=& m \exp\left(2 \int_{\tau_0}^{\tau} \epsilon(\tau)
	\mbox{d}\tau \right), \label{M:t} \\
\omega^2(\tau) &=& \omega_0^2 + \frac{\dot{e}^2(\tau)}{4m}, \\
F(\tau) &=& - \frac{e(\tau)}{m}  \frac{d\phi^0}{d\tau} (t(\tau),x(\tau)),
\end{eqnarray}
where $\epsilon(\tau) = e^2(\tau)/(4m)$.
Eq. (\ref{q'':F})  is the equation of motion of a damping harmonic 
oscillator subject to an external force $F(\tau)$.

Now let us find $B_Q(t)$ which satisfies
the commutation relation
\begin{eqnarray} \label{commutator}
[B_Q(\tau), B^\dagger_Q(\tau) ] = 1,  ~
[B_Q(\tau), b^{(+)}_\kappa]
	= [B_Q(\tau),b^{(+)\dagger}_\kappa ]=0, 
\end{eqnarray}
and satisfies the first order differential equation \footnote{
This equation was found in Ref.~\cite{kim}, where the authors
quantized the time-dependent-forced harmonic oscillator.}
\begin{eqnarray}
%\label{eqofmotion}
\frac{d}{d\tau}B_Q(\tau) =
	 -i \frac{\omega_I}{M(\tau)g_-(\tau)} B_Q(\tau) ,
\end{eqnarray}
whose solution is 
\begin{eqnarray}
%\label{}
B_Q(\tau) = e^{-i \Theta(\tau)} B_Q(\tau_0),
\end{eqnarray}
where $\displaystyle \Theta(\tau)\equiv
 \int d\tau\frac{\omega_I}{M(\tau)g_-(\tau)}$ and
$\omega_I^2 \equiv g_+g_- - g_0^2 $ is invariant
under time evolution.
The function $g_i$'s  are defined in Refs.~\cite{kim,jyji}
%especially in this paper, 
are given by
\begin{eqnarray}
%\label{}
g_-(\tau) \equiv f(\tau)f^*(\tau),
~g_0(\tau) \equiv -\frac{M(\tau)}{2}\dot{g}_-(\tau),~
g_+(\tau)\equiv M^2(\tau)\left|\dot{f}(\tau)\right|^2,
\end{eqnarray}
where $f(\tau)$ is a homogeneous solution of Eq. (\ref{q'':F}).
The explicit form of $B_Q(\tau)$ can be found by setting 
\begin{eqnarray}
%\label{}
B_Q(\tau) \equiv b(\tau) + e^{-i \Theta(\tau)} \beta(\tau),
\end{eqnarray}
and then 
by finding a differential equation which is satisfied with $\beta(\tau)$:
\begin{eqnarray}
\label{beta:F}
\frac{d}{d\tau}\beta(\tau) 
	= -i \sqrt{\frac{g_-(\tau)}{2 \omega_I}}M(\tau)F(\tau) .
\end{eqnarray}
Here $\displaystyle b(\tau)=\sqrt{\frac{g_+(\tau)}{2
	 \omega_I}}e^{i\xi(\tau)} q(\tau) + 
	i \sqrt{\frac{g_-(\tau)}{2 \omega_I}} \frac{M(\tau)}{m} p(\tau)$ 
is the annihilation operator of an oscillator~\cite{jyji}
and 
\begin{eqnarray}
%\label{}
\displaystyle \tan \xi(\tau) \equiv \frac{g_0(\tau)}{\omega_I} .
\end{eqnarray} 

By explicitly integrating Eq. (\ref{beta:F}) one get
\begin{eqnarray}
\label{beta:g}
\beta(\tau) =
	\beta_{Q0} + \sqrt{\frac{2 \omega_I}{g_-(\tau)}}
        \int_0^\infty d\Omega \Omega 
	\sqrt{2}\left[b^{(+)}_{\Omega +} g^*(-\Omega,\tau)
	- b^{(+) \dagger}_{\Omega +} g^*(\Omega,\tau) \right], 
\end{eqnarray}
where
$\beta_{Q0}$ is a constant operator which should be fixed by
the initial condition of $B_Q(\tau)$ 
and $g(\Omega,\tau)$ is
\begin{eqnarray}
\label{g:u}
g(\Omega,\tau) = \frac{\sqrt{g_{-}(\tau)}}{2 m \omega_I} 
	\int_{\tau_0}^\tau \mbox{d}\tau'
		\sqrt{g_-(\tau')} M(\tau') e(\tau') e^{-i \Theta(\tau')}
		u_\Omega (\tau',0).
\end{eqnarray}
Let us set the invariant operators to describe the initial
state of the oscillator at $\tau_0$: 
\begin{eqnarray}
\label{BQ:0}
B_Q(\tau_0)= a(\tau_0) = \sqrt{ \frac{m\omega_0}{2}} q(\tau_0) 
	+ i \sqrt{\frac{1}{2 m \omega_0}}p(\tau_0) .
\end{eqnarray}
From the requirement (\ref{BQ:0}) we determine
the unknown constant operator in Eq. (\ref{beta:g})
\begin{eqnarray}
\label{beta:B}
\beta_{Q0} &=& (1-c_1) e^{i\Theta(\tau)} 
	B_Q(\tau) -c_2 e^{-i \Theta(\tau)}
		B_Q^\dagger(\tau) , 
\end{eqnarray}
where 
\begin{eqnarray}
%\label{}
 c_1&=&
	\frac{1}{2}\sqrt{\frac{g_+(\tau_0)}{m\omega_0\omega_I}} \left[
	e^{i \xi(\tau_0)}+ m\omega_0 \sqrt{\frac{g_-(\tau_0)}{
		g_+(\tau_0)}} 
	\right]  , \\ 
c_2&=&\frac{1}{2}\sqrt{\frac{g_+(\tau_0)}{m\omega_0\omega_I}} \left[
	e^{i \xi(\tau_0)}- m\omega_0 \sqrt{\frac{g_-(\tau_0)}{
		g_+(\tau_0)}} 
	\right]  .
\end{eqnarray}
Now we finally find 
\begin{eqnarray}
\label{B_Q}
\bar{B}_Q(\tau) &=& \sqrt{\frac{m}{M(\tau)} }
	 e^{i\Theta} B_Q(\tau)  \nonumber \\ 
&=&  \alpha
    \sqrt{\frac{m g_+(\tau)}{2M(\tau)\omega_I}}q(\tau) 
	+  i\beta\sqrt{\frac{M(\tau)g_-(\tau)}{2m\omega_0}}p(\tau)   \\
 &+&  \sqrt{\frac{2m\omega_I}{M(\tau)g_-(\tau)}}
	\int d\Omega \Omega \sqrt{2}\left[
	b^{(+)}_{\Omega +}G_1(\Omega,\tau)  - 
	b_{\Omega +}^{(+) \dagger} 
	G_1(-\Omega,\tau) \right]    \nonumber
\end{eqnarray}
where $\displaystyle \alpha = c_1^* e^{i(\Theta(\tau) +\xi(\tau) )}+ 
	c_2e^{-i(\Theta +\xi)  }$,
$\beta =  c_1^* e^{i \Theta(\tau)} - c_2 e^{-i\Theta(\tau)} $, and
\begin{eqnarray}
%\label{}
G_1(\Omega,\tau)= c_1^*g^*(-\Omega,\tau)+ c_2 g(\Omega,\tau).   
\end{eqnarray}

By inverting Eq. (\ref{B_Q}) one get the time evolution of the 
oscillator
\begin{eqnarray}\label{q:0A}
q(\tau) &=& q_O(\tau)+ q_F(\tau) \\
	&=& \sqrt{ \frac{g_-(\tau)}{2 \omega_I}} 
	\left(c_1+ c_2 ^* e^{2i \Theta(\tau)} \right)B_Q(\tau) + 
	 B_F(\tau)
	+{\rm  H. C.} , 
	 \nonumber
\end{eqnarray}
where $q_F(\tau)= B_F(\tau) + B_F^\dagger(\tau)$ and
\begin{eqnarray}
B_F(\tau) =  \int_0^{\infty} \mbox{d}\Omega  \Omega 
	\sqrt{2} G_-(\Omega,\tau) b^{(+)}_{\Omega +}.
\end{eqnarray}
In this equation,
\begin{eqnarray}
%\label{}
G_-(\Omega,\tau)= e^{i \Theta(\tau)}
	g(\Omega,\tau)- e^{-i \Theta(\tau)} g^*(-\Omega,\tau)
\end{eqnarray} 
is  a classical solution of the forced harmonic oscillator 
equation (\ref{q'':F})
with $\displaystyle F(\tau)= -\frac{ie(\tau)}{m} 
u_\Omega^{(+)}(\tau,0)$ and
its initial conditions are 
\begin{eqnarray} \label{condition1}
G_-(\Omega,\tau_0)=0 , ~   \dot{G}_-(\Omega,\tau_0) = 0.
\end{eqnarray}
If the coupling does not vary,  
$\bar{B}_Q$ in Eq. (\ref{B_Q}) asymptotically vanishes 
because $M(\tau)$ increase exponentially.
So, the dynamics of the oscillator is completely determined 
by the field asymptotically.

The commutation relations (\ref{commutator}) allow 
the natural definition of the Fock space  as eigenstates
of $B_Q^\dagger B_Q$, and $b^{(\sigma)\dagger}_k 
b^{(\sigma)}_k$. The eigenvalues of these operators
are invariant under time evolution of the system, so
we usually call them invariants.
%can define Fock space as an eigenstate of number operators:
%\begin{eqnarray}
%\label{invariant}
%I(\tau) = B_Q^\dagger(\tau)B_Q(\tau) , ~ 
%I_\kappa(\tau)=b^{(+)^\dagger}_\kappa b^{(+)}_\kappa ,
%\end{eqnarray} 
The ground state $\left|0\right>_R$ is 
\begin{eqnarray} \label{0:b}
b^{(\sigma)}_\kappa \left|0\right>_R = B_Q(\tau)\left|0 \right>_R =0 .
\end{eqnarray}
In fact, $\left|0\right>_R$ is the Rindler vacuum state at
$\tau_0$ because $B_Q(\tau)$ are simply proportional to
$a(\tau_0)$ and $b_\kappa^{(\sigma)}$ are invariant on time.

By inserting Eqs. (\ref{phi:uv}) and (\ref{q:0A}) into (\ref{B})
we obtain the relations between the two sets of operators. 
$B^{(-)}_\kappa$ is not affected by the presence of the oscillator
and $B^{(+)}_\kappa$ is given by
\begin{eqnarray}
\label{A:a}
B_\kappa^{(+)} (\tau) &=& b_\kappa^{(+)} 
	+ \Omega \int^\tau_{\tau_0} 
		u_\kappa^{(+)*}(\tau',0) 
	e(\tau')q(\tau') d\tau' \nonumber \\
	&=& b_\kappa^{(+)} + 
		\Omega \int^\tau_{\tau_0} 
		u_\kappa^{(+) *} (\tau',0) 
		e(\tau') q_O(\tau') d\tau'   \\ 
	&+& \Omega \int d\Omega' \Omega' 
	\bar{G}(\Omega,\Omega',t) \sqrt{2}b_{\Omega' +}^{(+)}
	 +  \Omega \int d\Omega' \Omega' 
	 \bar{G}^*(-\Omega,\Omega',\tau) 
	\sqrt{2}b^{(+)\dagger}_{\Omega' +} , \nonumber
\end{eqnarray}
where 
\begin{eqnarray}
\label{barG}
\bar{G}(\Omega,\Omega',\tau) = \int^\tau d\tau'e(\tau') u^{(+)}_\Omega(\tau',0) G_-(\Omega',\tau').
\end{eqnarray} 
From Eq. (\ref{B_Q}) we get the instantaneous annihilation 
operator of the oscillator 
\begin{eqnarray}
%\label{}
a(\tau) &=&\sqrt{\frac{mg_-(\tau) \omega_0}{4\omega_I}} \left[  
	\left\{ \left(c_1 +c_2^*e^{2i\Theta}\right) B_Q(\tau) 
		+H.C. \right\}  \right. \nonumber \\
  &+& \left. \frac{\omega_I}{M(\tau)g_-(\tau) \omega_0} 
	\left\{ (c_1-c_2^*e^{2i\Theta} ) B_Q(\tau)
	+ H.C.  \right\} \right] \\
&+&\sqrt{\frac{m \omega_0}{2} }
\left[ \int 
	d\Omega \Omega
	\left\{ b_{\Omega+}^{(+)}G_-(\Omega,\tau)+ 
	b_{\Omega+}^{(+)\dagger} 
	G_-^*(\Omega,\tau)  \right\} \right. \nonumber \\
    &-& \left. \frac{\omega_I}{M(\tau)g_-(\tau)\omega_0} 
		\int d\Omega \Omega 
	\left\{ b_{\Omega +}^{(+)}G_+(\Omega,\tau)
	- b_{\Omega +}^{(+)\dagger} G_+^*(\Omega,\tau)   \right\}
	\right] ,  \nonumber 
\end{eqnarray}
where $G_+(\Omega,\tau) = e^{i\Theta(\tau)}g(\Omega,\tau) +
	e^{-i\Theta(\tau)} g^*(-\Omega,\tau)$.

%%%%%%%%%%%%%%%%%%%%%%%%%%%%%%%%%%
%
%
%
%
%%%%%%%%%%%%%%%%%%%%%%%%%%%%%%%%%%
%%%%%%%%%%%%%%%%%%%%%%%%%%%%%%%%%%%%%%%
\section{Appendix B: Operator relations in constant coupling}

We are interested in the final state of the system,
which can be obtained by taking the asymptotic limit  
of a constant coupling.  In this paper, 
the asymptotic limit is to take  $\tau_0 \rightarrow -\infty$. 
From now on, we set $e(\tau)=e$.

The homogeneous solution of the
oscillator decays to zero:
\begin{eqnarray}
%\label{}
q_O(\tau) = \left[ q(\tau_0)\frac{\omega_0}{\omega_I}\cos\{\omega_I(\tau-\tau_0)-\theta\} +
	 \frac{p(\tau_0)}{m \omega_I} \sin \omega_I(\tau-\tau_0)
 	 \right] e^{-\epsilon(\tau-\tau_0)}.
\end{eqnarray}
The inhomogeneous solution is
\begin{eqnarray} \label{G:const}
&&G_-(\Omega,\tau) =
i e\chi(\Omega)	u_\Omega^{(+)}(\tau) \\
&&+ i e \chi(\Omega) u^{(+)}_\Omega(\tau_0)
	e^{-\epsilon(\tau-\tau_0)} \left[ - \cos[\Omega_I(\tau-\tau_0) -\theta]   -
	i \frac{\Omega}{\omega_I}\sin\omega_I(\tau-\tau_0) \right]  \nonumber ,
\end{eqnarray}
where 
\begin{eqnarray}
\label{chi}
\chi(\Omega) = \frac{1}{m[\omega_0^2- \Omega^2-2i \epsilon \Omega ]} .
\end{eqnarray}
The first term of Eq. (\ref{G:const}) vanishes asymptotically.  
From now on, we ignore terms which vanishes for 
the limit $\tau_0\rightarrow -\infty$. 
The asymptotic form of $a(\tau)$ is 
\begin{eqnarray}
\label{a:b}
a(\tau)&=&
	 i e \sqrt{m\omega_0}\left\{ \frac{\omega_0
	+i\epsilon}{\omega_0} \int d \Omega \Omega
	\left[ u_\Omega^{(+)}(\tau,0) \chi(\Omega)
	b^{(+)}_{\Omega +} \right. \right. \\
	&&\left. \left.- u_\Omega^{(+)*}(\tau,0) \chi^*(\Omega) 
	 b_{\Omega +}^{(+)\dagger} \right]    \right. \nonumber  \\
& - & \left. \frac{1}{\omega_0}\int d\Omega \Omega \left[   
	b_{\Omega +}^{(+)} (\Omega + i\epsilon) \chi(\Omega)  u_\Omega(\tau)
	+ b_{\Omega +}^{(+)\dagger} (\Omega-i\epsilon) \chi^*(\Omega)
	u_\Omega^*(\tau,0) 
\right]  \right\} \nonumber
\end{eqnarray}

Now we are ready to get the asymptotic
relation between $B_\kappa^{(\sigma)}$ and 
$b_\kappa^{(\sigma)}$.  The expansion 
coefficients (\ref{barG}) are
\begin{eqnarray}
\label{Gbar}
\bar{G}(\Omega,\Omega',\tau)&=& 
 		 \frac{i e^2 \chi(\Omega)}{\sqrt{2 \Omega}} \delta(\Omega-\Omega') 
	+O(1/\tau_0),
\end{eqnarray}
and $\Omega  \int d\tau' u_k^{(+)*}(\tau',0) q_O(\tau') \sim O(1/\tau_0)$. 
In these calculations, we use the formula:
$\displaystyle \lim_{x\rightarrow \infty} e^{i \alpha x} 
	\simeq \frac{\delta(\alpha)}{x}$.
From Eq. (\ref{Gbar}) we get 
\begin{eqnarray}
\label{B:b}
B_\kappa^{(\sigma)}(\tau) = b_\kappa^{(\sigma)} + 
	 2 i m \epsilon \Omega \chi(\Omega) 
	\sqrt{2}b^{(+)}_{\Omega +} \delta_{\sigma +}.
\end{eqnarray}
One can easily verify the commutation relation (\ref{comm}) using the
Kramer-Kroig type relation $\chi(\Omega) -\chi^*(\Omega)
=4im \epsilon \Omega |\chi(\Omega)|^2 $.  This transformation 
is non-singular in a real $\kappa$ axis so the inverse
can be obtained easily:
\begin{eqnarray}
\label{b:B}
b^{(+)}_\kappa = \frac{B^{(+)}_\kappa(\tau)
	+ 2im\epsilon \Omega \chi(\Omega)
	 \left[B^{(+)}_\kappa(\tau)-B^{(+)}_{-\kappa}(\tau)\right] 
	}{1+4im\epsilon \Omega \chi(\Omega)} .
\end{eqnarray}

%%%%%%%%%%%%%%%%%%%%%%%%%%%%%%%%%%%%%%

\end{document}